\documentclass[prd,twocolumn,showpacs,showkeys,preprintnumbers]{revtex4}
\usepackage{hyperref,amssymb,amsmath,mathrsfs,bm,graphicx}
\usepackage{color}
\begin{document}
\title{On the equivalence of nonadiabatic fluids}
\date{\today}
\author{W. Barreto}
\email{wbarreto@ula.ve}
\affiliation{Centro de F\'\i sica Fundamental, Facultad de Ciencias, Universidad de Los Andes, M\'erida, Venezuela\footnote{On sabbatical leave while beginning this work.}}
\begin{abstract}
{Here we show how an anisotropic fluid in the diffusion limit can be equivalent to an isotropic fluid in the streaming out limit, in spherical symmetry. For a particular equation of state this equivalence is total, from one fluid we can obtain the other and vice versa. A numerical master model is presented, based on a generic equation of state, in which only quantitative differences are displayed between both non--adiabatic fluids. From a deeper view, other difference between fluids is shown as an {\it asymmetry} that can be overcome if we consider the appropriate initial--boundary conditions. Equivalence in this context can be considered as a first order method of approximation to study dissipative fluids.} 

\end{abstract}
\maketitle

\section{Introduction}

Emission of massless particles is an important process when studying evolution
of massive stars. 
If the fluid is nonadiabatic two transport mechanisms are recognized as extreme and opposite, namely, the streaming out and the diffusion limits. In the event of gravitational collapse these two transport regimes alternate, playing different roles depending on the configuration and physical processes  \cite{ks79}, \cite{l88}, \cite{hd97}. 

It is well known that different energy--momentum tensors can lead to the same spacetime \cite{tt73}--\cite{ci85}. Different kinds of physical phenomena may take place to give rise to local anisotropy \cite{l33}. The influence of anisotropy in self--gravitating systems has been studied mostly under static conditions (see \cite{hs97} and references therein). 
In \cite{gdmost04} a general study of spherically symmetric dissipative anisotropic fluids was reported, with emphasis on the relationship between the Weyl tensor, the shear tensor, the anisotropy of the pressure and the density inhomogeneity. 
For instance, shear viscosity can be considered as a special case of anisotropy \cite{br92}--\cite{ct94}, \cite{prrb10}, with no contribution to energy density and energy flux as heat flow. Bulk viscosity does not change the degree of anisotropy. Electrically charged fluids also can be seen as anisotropic fluids with no contribution to heat flow \cite{brrs07}, \cite{rbrp10}. 

Many years ago the equivalence between radiating perfect fluids
and anisotropic fluids was shown \cite{l80}, which also applies under certain conditions 
to multicomponent fluids \cite{la86}, \cite{p87}. 
Recently it was proposed that an anisotropic fluid in the diffusive limit is, in some sense, the most fundamental model \cite{i10} because it can absorb the addition of shear viscosity, electric charge and null fluids. In fact, the free streaming process generates heat flow, radial pressure (nonisotropically) and energy density. In consequence the resulting fluid is anisotropic.

We show here how in practice an isotropic fluid in the streaming out limit
can be equivalent to an anisotropic fluid under a diffusion transport mechanism. 
We argue that an isotropic fluid with free streaming is as fundamental as an
anisotropic fluid with heat flow. We illustrate
our claim numerically by means of a master model. To do that we use the seminumeric
method known as the postquasistatic approach (see \cite{hbds02} and references therein),
 which fits well for this paper purposes. 
We do not know of any other hydro solver in numerical relativity that considers a radiating fluid in either regime, streaming out or diffusion, isotropic or anisotropic, in spherical symmetry.
 
The equivalence between nonadiabatic fluids as presented here can be used to study different situations in the gravitational collapse with the same fluid, but this is not always possible. The equation of state seems to be crucial as well as the initial conditions, the exterior spacetime and the junction conditions.
The equivalence has been used as a method to solve the field equations, but not
to extract physics. This situation is quite similar to a massless scalar field viewed as a fluid. The asymptotic conditions near the center of symmetry, at the boundary surface and near infinity are different for the scalar field, in comparison with a bounded source of
matter \cite{bcb09}, \cite{bcb10}. 

{The remaining parts of this work are organized as follows: In Section II, we show how a local comoving observer can handle the equivalence between non-adiabatic fluids and how the field equations can be recast. In Section III, we give some examples using a master numerical model. In section IV, we revisit the results in the light of previous ones and conclude with some remarks  about the limits and scope of the equivalence of nonadiabatic fluids.}
\section{Equivalent radiating fluids}
We can clearly show at once the aforementioned
equivalence between nonadiabatic fluids using
Bondian observers (comoving in a locally Minkowskian spacetime) to describe how physics
comes out from hydrodynamics. The covariant energy--momentum tensor for a radiating perfect fluid as seen by a comoving observer with respect to a local one with
radial velocity $\omega$ is
\begin{equation}
\left(
\begin{array}{cccc}
\hat\rho + \epsilon & -\hat\epsilon & 0 & 0 \\
-\hat\epsilon & \hat p + \hat\epsilon & 0 & 0 \\
0 & 0 & \hat p & 0 \\
0 & 0 & 0 & \hat p 
\end{array}
\right), \label{TEM1}
\end{equation}
representing a radiating isotropic fluid of energy density $\hat\rho$, pressure $\hat p$, 
and energy density $\hat\epsilon$ traveling in the radial direction. 
If we define
\begin{subequations}
\begin{eqnarray}
\rho&=&\hat\rho + \epsilon,\label{d}\\
\hat q&=&\hat\epsilon,\label{q}\\
p_r&=& \hat p + \hat\epsilon, \label{pr}\\
p_t&=&\hat p, \label{pt}
\end{eqnarray} 
\end{subequations}
the fluid can be seen as
%\equiv
\begin{equation}
\left(
\begin{array}{cccc}
\rho & -\hat q & 0 & 0 \\
-\hat q & p_r & 0 & 0 \\
0 & 0 & p_t & 0 \\
0 & 0 & 0 &  p_t 
\end{array}
\right), \label{TEM2}
\end{equation}
which can be interpreted as an anisotropic fluid with effectives energy
density $\rho$, radial pressure $p_r$, tangential pressure $p_t$ and heat flow
in the radial direction $q$.
Observe that the degree of anisotropy is
\begin{equation}
\Delta\equiv p_t-p_r=-\hat\epsilon,\label{ieos}
\end{equation}
which in turn is seen as an equation of state.
As a matter of fact, any comoving observer sees such equivalence, it does not depend on 
the system of coordinates. No rotation about the timelike two--plane is required \cite{l80}, \cite{la86}, \cite{i09} because they are implicit in (\ref{TEM1}) and (\ref{TEM2}). It is manifest that the covariant energy--momentum tensor is invariant. 

We have to write the field equations in one limit to show explicitly and formally how we
can get the equivalence simply recasting them.
Using Bondi's metric in spherical form \cite{b64}
\begin{equation}
ds^2=e^{2\beta}\Bigl(\frac{V}{r}du^2+2dudr\Bigr)
     -r^2(d\theta^2+\sin\theta^2 d\phi^2), \label{bm}
\end{equation}
where $\beta=\beta(u,r)$ and $V=V(u,r)$.
The field equations for an isotropic fluid in the free streaming approximation can be written as
\cite{hjr80}

\begin{widetext}
\begin{eqnarray}
\frac{\hat\rho+\omega^2 \hat p}{1-\omega^2}+\hat\epsilon\frac{1+\omega}{1-\omega} =
\frac{1}{4\pi r(r-2\tilde m)}[-\tilde m_{,u}e^{-2\beta}
+ (r-2\tilde m)\tilde m_{,r}/r], \label{e1}
\end{eqnarray}

\begin{equation}
\frac{\hat\rho-\omega \hat p}{1+\omega}
=\frac{\tilde m_{,r}}{4\pi r^2},
\label{e2}
\end{equation}

\begin{equation}
\frac{1-\omega}{1+\omega}(\hat\rho+\hat p) =
 \frac{1-2\tilde m/r}{2\pi r}\beta_{,r},\label{e3}
\end{equation}
\begin{eqnarray}
\hat p= -\frac{1}{4\pi}\beta_{,ur}e^{-2\beta}
+ \frac{1}{8\pi}(1-2\tilde m/r) (2\beta_{,rr} + 4\beta^2_{,r}- \beta_{,r}/r)
+ \frac{1}{8\pi r}[3\beta_{,r}(1-2\tilde m_{,r})-
\tilde m_{,rr}],\label{e4}
\end{eqnarray}
\end{widetext}
where a comma denotes partial differentiation with respect to any coordinate,
and the mass aspect $\tilde m$ is defined by means of
\begin{equation}
V=e^{2\beta}(r-2\tilde m).
\end{equation}
As expected, we can write the LHS of the field equations in the following way, only rearranging terms 
\begin{eqnarray}
S\equiv\frac{\hat\rho+\omega^2 \hat p}{1-\omega^2}+\hat\epsilon\frac{1+\omega}{1-\omega}\equiv\frac{\rho +\omega^2 p_r}{1-\omega^2}+\frac{2\hat q\omega}{1-\omega^2},\label{ea}
\end{eqnarray}
\begin{eqnarray}
\tilde\rho\equiv\frac{\hat\rho-\omega \hat p}{1+\omega}
\equiv\frac{\rho-\omega  p_r}{1+\omega}-\hat q \frac{1-\omega}{1+\omega},
\label{eb}
\end{eqnarray}
\begin{eqnarray}
\tilde\rho+\tilde p\equiv\frac{1-\omega}{1+\omega}(\hat\rho+\hat p) \equiv\frac{1-\omega}{1+\omega}(\rho+p_r) -2\hat q \frac{1-\omega}{1+\omega} ,\label{ec}
\end{eqnarray}
where $\tilde\rho$ and $\tilde p$ are the so--called effective variables. In standard numerical relativity $S$ and $\tilde\rho$ are the conservative variables, and $\tilde p$ the flux variable. 
Now (\ref{e1})--(\ref{e4}) are the field equations for an anisotropic fluid in the diffusion limit with the LHS formed with (\ref{ea})--(\ref{ec}), (\ref{pt}) and the additional equation of
state (\ref{ieos}).
In this sense it is said that an isotropic fluid in the free--streaming limit is equivalent to an
anisotropic fluid in the diffusion limit. If the fluid is described {\it ab initio} as anisotropic in the diffusion limit and the equation of state to take into account the degree of
anisotropy is given by (\ref{ieos}), then we get necessarily an isotropic fluid with free streaming.
We shall explore a generic situation, that is, with any other equation of state. In that case: 
does the equivalence between these two nonadiabatic fluids hold? 

{In order to gain physical insight we write the field equation (\ref{e4}) as the generalized equation for hydrostatic support of Tolman--Oppenheimer--Volkoff \cite{hjr80},
\cite{chew82}
\begin{equation}
\tilde p_{,r}+\frac{(\tilde p + \tilde\rho)}{r(r-2\tilde m)}=e^{-2\beta}\left(\frac{\tilde p + \tilde\rho}{1-2\tilde m/r}\right)_{,u}+\frac{2}{r}(p_t-\tilde p).\label{TOV}
\end{equation}}

One way to illustrate our view point is using a method which represents the departure from
equilibrium in the gravitational collapse problem \cite{hjr80}. That method reported many years ago could be used as a testbed for numerical relativity \cite{b09}. It as been interpreted as a postquasistatic approximation \cite{hbds02}. The procedure is based on the cornerstone paper of Bondi to deal with radiating matter \cite{b64}.

For details about the method used to solve the equations, matching conditions 
and its connection with standard numerical relativity see \cite{hjr80}, \cite{hbds02}, \cite{b09}. Here we only write the appropriate boundary condition \cite{hje87}
\begin{equation}
[p_r]_a=\hat q_a,
\end{equation}
which is a consequence of the Darmois--Lichnerowicz conditions, that is, the continuity of the first and second differential forms:
\begin{equation}
m=\mathcal{M}(u),
\end{equation}
\begin{equation}
\beta_a=0,
\end{equation}
\begin{equation}
\left[-\beta_{,u}e^{-2\beta}+(1-2\tilde m/r)\beta_{,r} - \frac{\tilde m_{,r}}{2r}\right]_{r=a}=0,
\end{equation}
where the subscript $a$ indicates that the quantity is evaluated at the surface $r=a(u)$,
$\tilde m_a=m$ 
and $\mathcal{M}$ corresponds to exterior Vaidya spacetime
\begin{equation}
ds^2=\left(1-\frac{2\mathcal{M}}{r}\right)du^2+2dudr-r^2(d\theta^2 + \sin\theta^2 d\phi^2).
\end{equation}
Clearly in the case of free streaming the boundary condition reads 
\begin{equation}\hat p_a=0.\end{equation}
\begin{figure}[htbp!]
\begin{center}
\scalebox{0.45}{\includegraphics[angle=0]{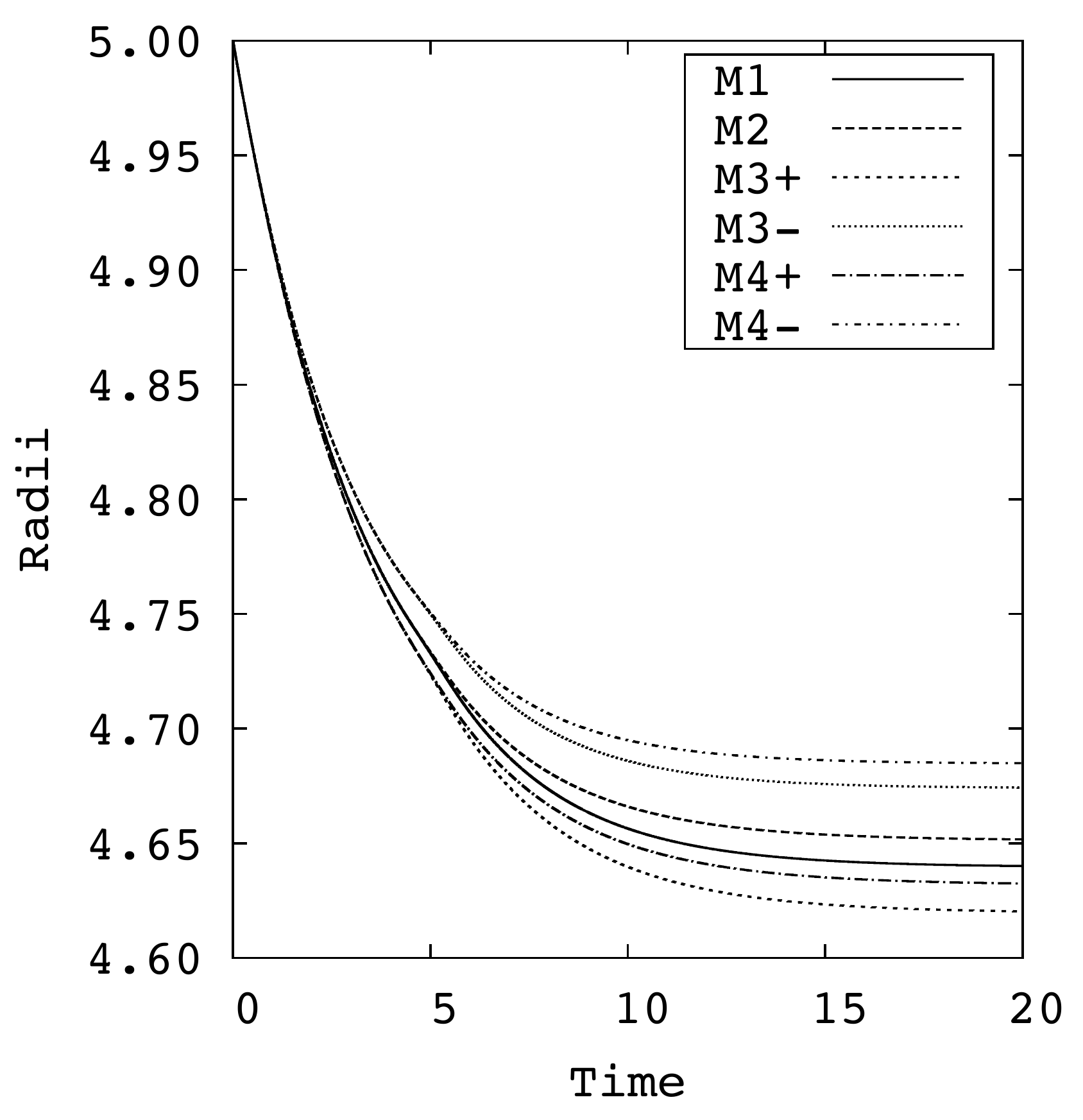}}
\caption{Evolution of the radius $a$ for different models: M1; M2; M3- with $h=0.33$; M3+ with $h=1.33$; M4- with $h=0.33$; M4+ with $h=1.33$; all radiating the same fraction of the initial total mass ($1\%$). The initial conditions are: $A(0)=5;\,\,F(0)=0.6;\,\, \Omega(0)=0.8333$. Model M4$^+$ may replicate model M1 if the anisotropy parameter $h\approx 1.2$.}
\end{center}
\label{fig:dens}
\end{figure}
\begin{figure}[htbp!]
\begin{center}
\scalebox{0.45}{\includegraphics[angle=0]{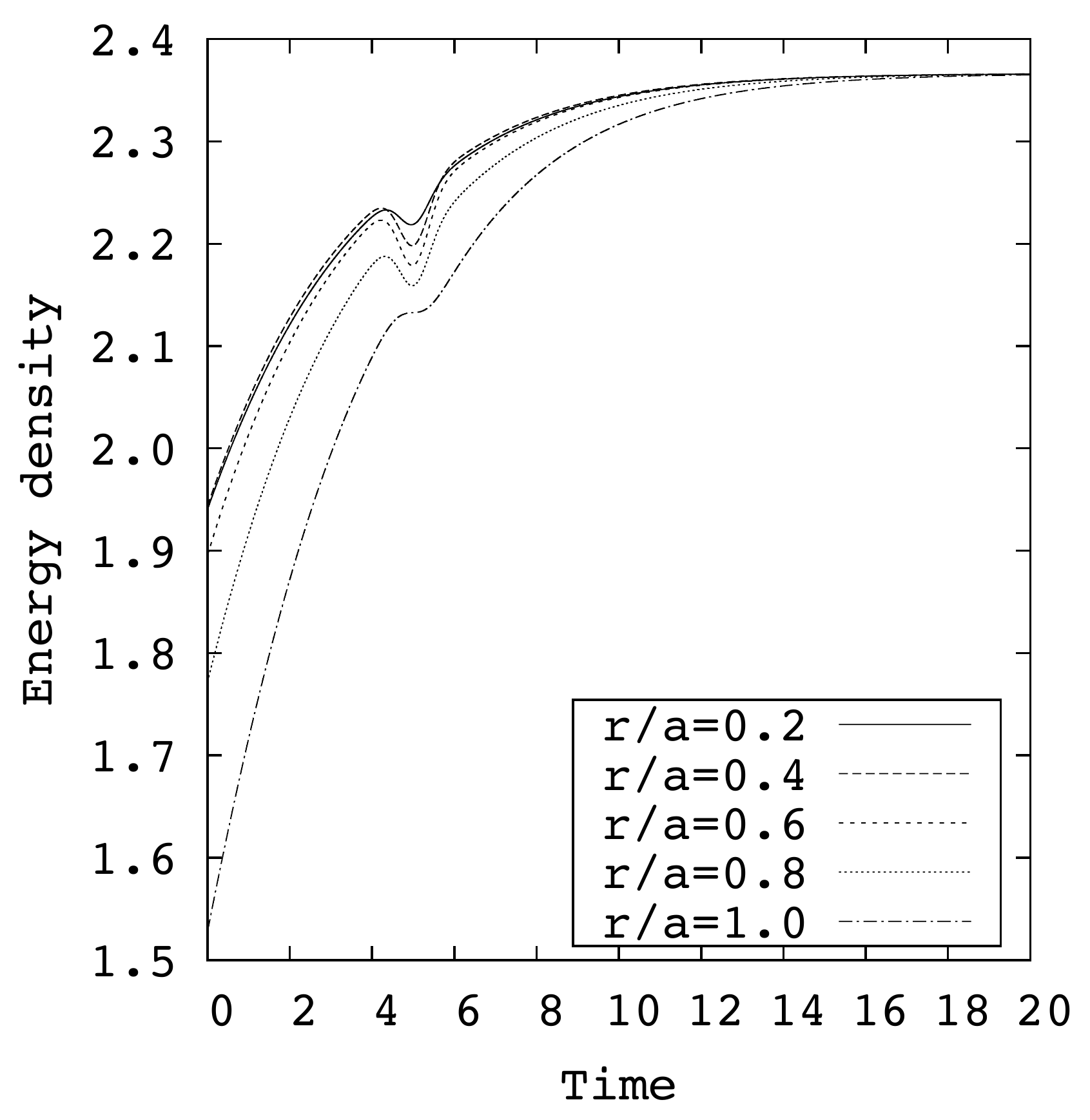}}
\caption{Evolution of the energy density $\hat\rho$ (multiplied by $10^3$) for the streaming out limit model (M1) and initial conditions $A(0)=5;\,\,F(0)=0.6;\,\, \Omega(0)=0.8333$.}
\end{center}
\label{fig:dens}
\end{figure}

To numerically study the equivalence, we shall use a Schwarzschild--like model with intrinsic anisotropy based on the following generic--equation--of--state (GEoS):
\begin{equation}\Delta=C\frac{(\tilde\rho + \tilde p)}{(r-2\tilde m)}(4\pi r^3 \tilde p + \tilde m),\label{CEO}\end{equation}
where $C$ is a constant.  This GEoS has been used in the past for a number of physical situations which represent phase transition, viscous and electrically charged fluids, \cite{chew82}, \cite{hn89}, \cite{br92}, \cite{b93}, \cite{prrb10}, \cite{rbrp10}.

\section{Master model}
{Here we describe a master model, which includes all the possible configurations of interest for our purposes, based on the models described in previous investigations \cite{hjr80}, \cite{hje87}, \cite{chew82}, \cite{br92}.}
Following the postquasistatic protocol, we build a Schwarzs\-child--like master model which corresponds to an incompressible anisotropic/isotropic fluid as the static ``seed'' interior solution. 
Thus, the effective energy density and the effective pressure are 
\begin{equation}
\tilde\rho=f(u)=\frac{3m}{4\pi a^3},
\end{equation}
\begin{equation}
\tilde p={\tilde\rho}\left\{\frac{(1-3\omega_a)\xi-(1-\omega_a)\xi_a}
{3(1-\omega_a)\xi_a-(1-3\omega_a)\xi}\right\},
\end{equation}
where
$$\xi=\left[1-\frac{2m}{a}\left(\frac{r}{a}\right)^2\right]^{h/2},$$
and
$$h=1-2C.$$
With these effective variables and integrating (\ref{e2}) and (\ref{e3}) we obtain
\begin{equation}
\tilde m={m}\left(\frac{r}{a}\right)^3
\end{equation}
\begin{equation}
\beta=\frac{1}{2h}\ln\left\{(1-\omega_a)\left[\left(\frac{3}{2}\frac{\xi_a}{\xi}-\frac{1}{2}\right)\right]+\omega_a\right\}
\end{equation}
The system of equations at the surface is
\begin{subequations}
\begin{eqnarray}
\dot A &=& F(\Omega-1),\label{se1}\\
\dot F &=&\frac{2L+(1-F)\dot A}{A},\label{se2}\\
(1-F)\frac{\dot\Omega}{\Omega} + \frac{\dot F}{F} &=& G \label{MMc}
\end{eqnarray}
where 
$$A=\frac{a}{m(0)}, \;\;\; \tau=\frac{u}{m(0)}, \;\;\; M=\frac{m}{m(0)},$$
$$F=1-\frac{2M}{A},$$
$$\Omega=\frac{1}{1-\omega_a},$$
\begin{eqnarray}
G=&-&\frac{3(1-F)^2(\Omega-1)}{2A\Omega}(2\Omega-1)\nonumber\\
&-&\frac{3(1-F)^2(\Omega-1)}{2A\Omega}(3-2\Omega)(1-h)\nonumber\\
&+&\frac{4L\Omega}{3A(2\Omega-1)}(1-\ell),\nonumber
\end{eqnarray}
\end{subequations}
overdot denotes the total derivative with respect to $\tau$, $L$ is the luminosity as a function of time given in some convenient way. {Thus, the master model
summarizes in one equation, (\ref{MMc}), four previously reported models of the same
type (Schwarzschild--like). The whole model is represented by the set of equations (26) and allows to control the transport mechanism and anisotropy using only two parameters ($h$ and $l$).} $h=1$ represents isotropy, $h<1$ implies $p_t > p_r$, and $h>1$ means $p_t < p_r$.
On the other hand, $\ell=1$ is for the streaming out limit and $\ell=0$ for 
the diffusion limit. Therefore, this master model contains all possible situations,
that is, isotropy in the streaming out limit (M1: $h=1$; $\ell=1$); isotropy in the diffusion limit (M2: $h=1$; $\ell=0$);
anisotropy in the streaming out limit (M3$\pm$: $h\ne1$; $\ell=1$); anisotropy in the diffusion limit (M4$\pm$: $h\ne1$; $\ell=0$). Equivalent models come from the corresponding master models depending on the equation of state used, for instance, E1 comes from M1 using (\ref{ieos}); E2 comes from M4+ which in turn is built using (\ref{CEO}). In other words, equivalent models are obtained with no additional calculations but just singling out the source model data.
\begin{figure}[htbp!]
\begin{center}
\scalebox{0.45}{\includegraphics[angle=0]{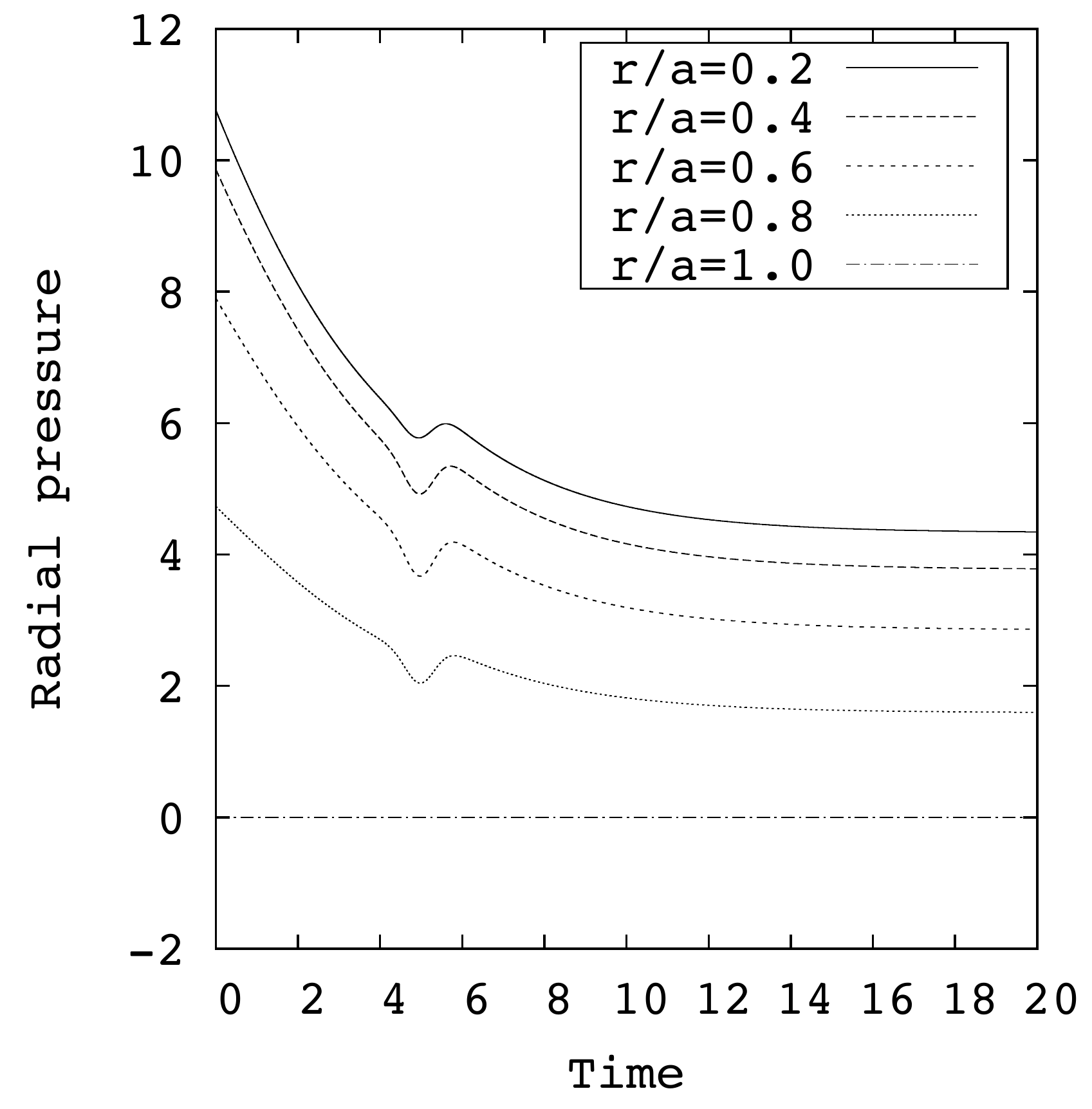}}
\caption{Evolution of the radial pressure $\hat p$ (multiplied by $10^4$) for the streaming out limit model (M1) and initial conditions $A(0)=5;\,\,F(0)=0.6;\,\, \Omega(0)=0.8333$.}
\end{center}
\label{fig:pres}
\end{figure}
\begin{figure}[htbp!]
\begin{center}
\scalebox{0.45}{\includegraphics[angle=0]{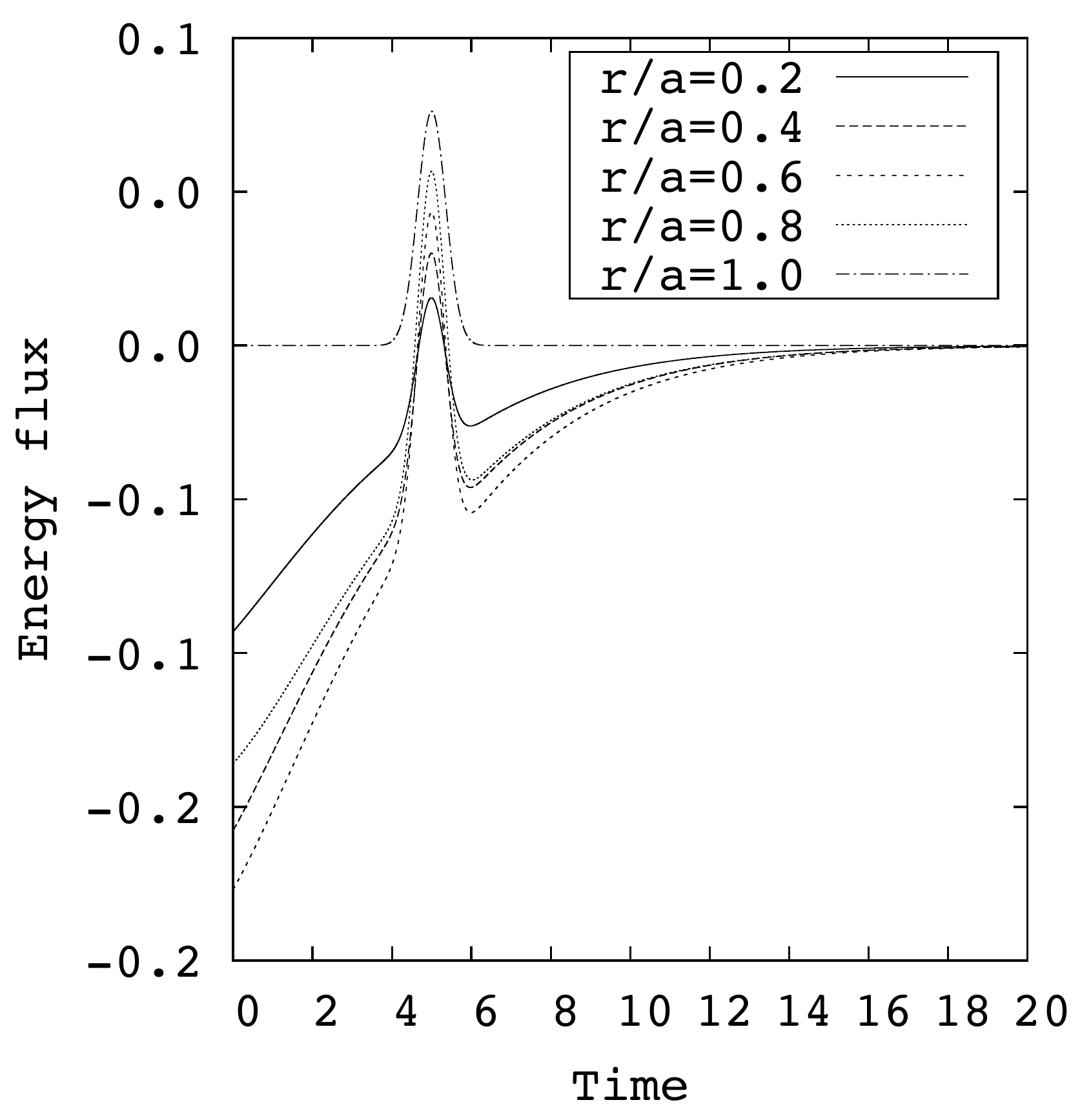}}
\caption{Evolution of the energy flux $\hat\epsilon$ (multiplied by $10^3$) for the streaming out limit model (M1) and initial conditions $A(0)=5;\,\,F(0)=0.6;\,\, \Omega(0)=0.8333$.}
\end{center}
\label{fig:flux}
\end{figure}
\begin{figure}[htbp!]
\begin{center}
\scalebox{0.45}{\includegraphics[angle=0]{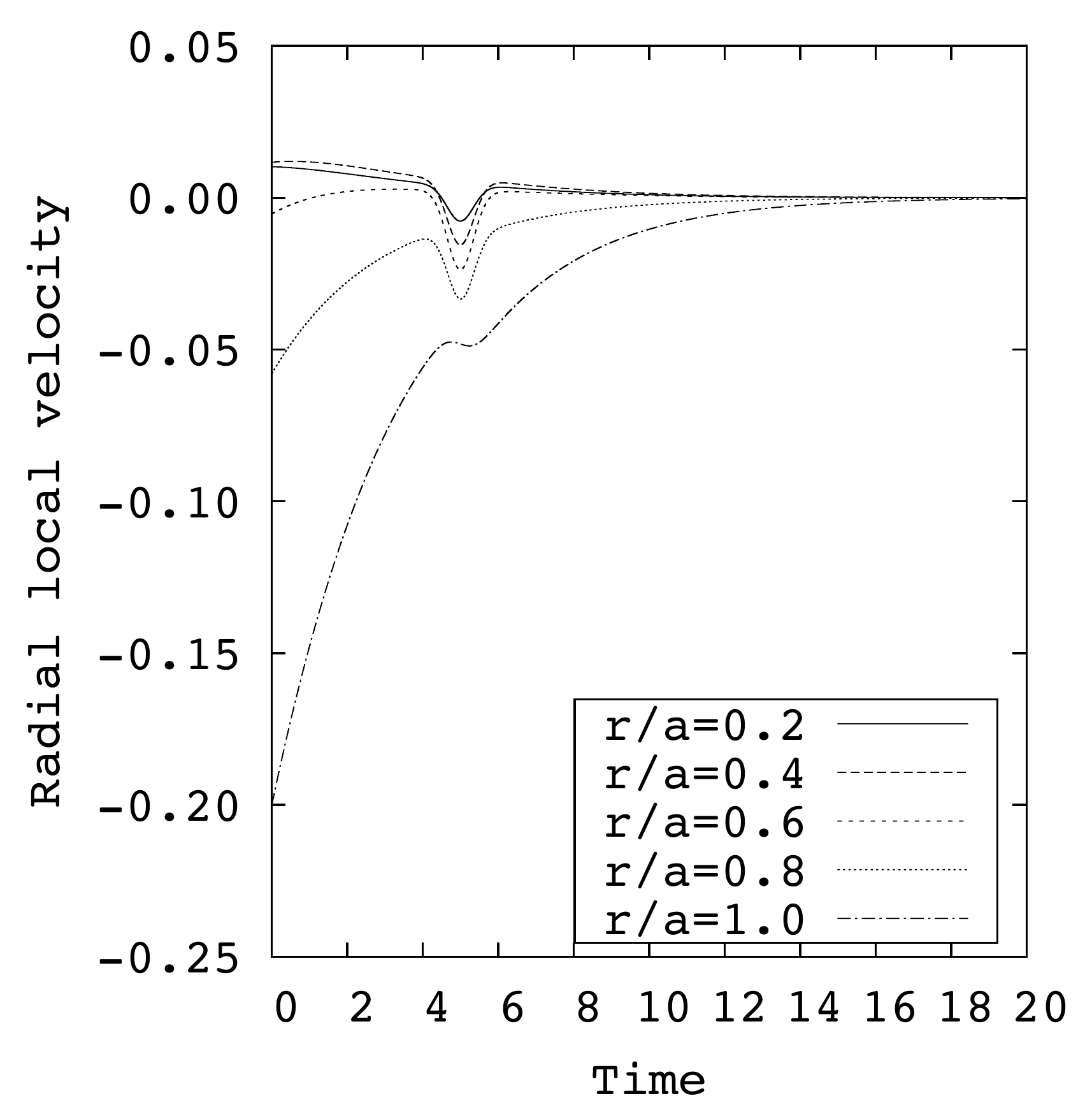}}
\caption{Evolution of the radial local velocity $\omega$ for the streaming out limit model (M1) and initial conditions $A(0)=5;\,\,F(0)=0.6;\,\, \Omega(0)=0.8333$.}
\end{center}
\label{fig:velocity}
\end{figure}
\begin{figure}[htbp!]
\begin{center}
\scalebox{0.45}{\includegraphics[angle=0]{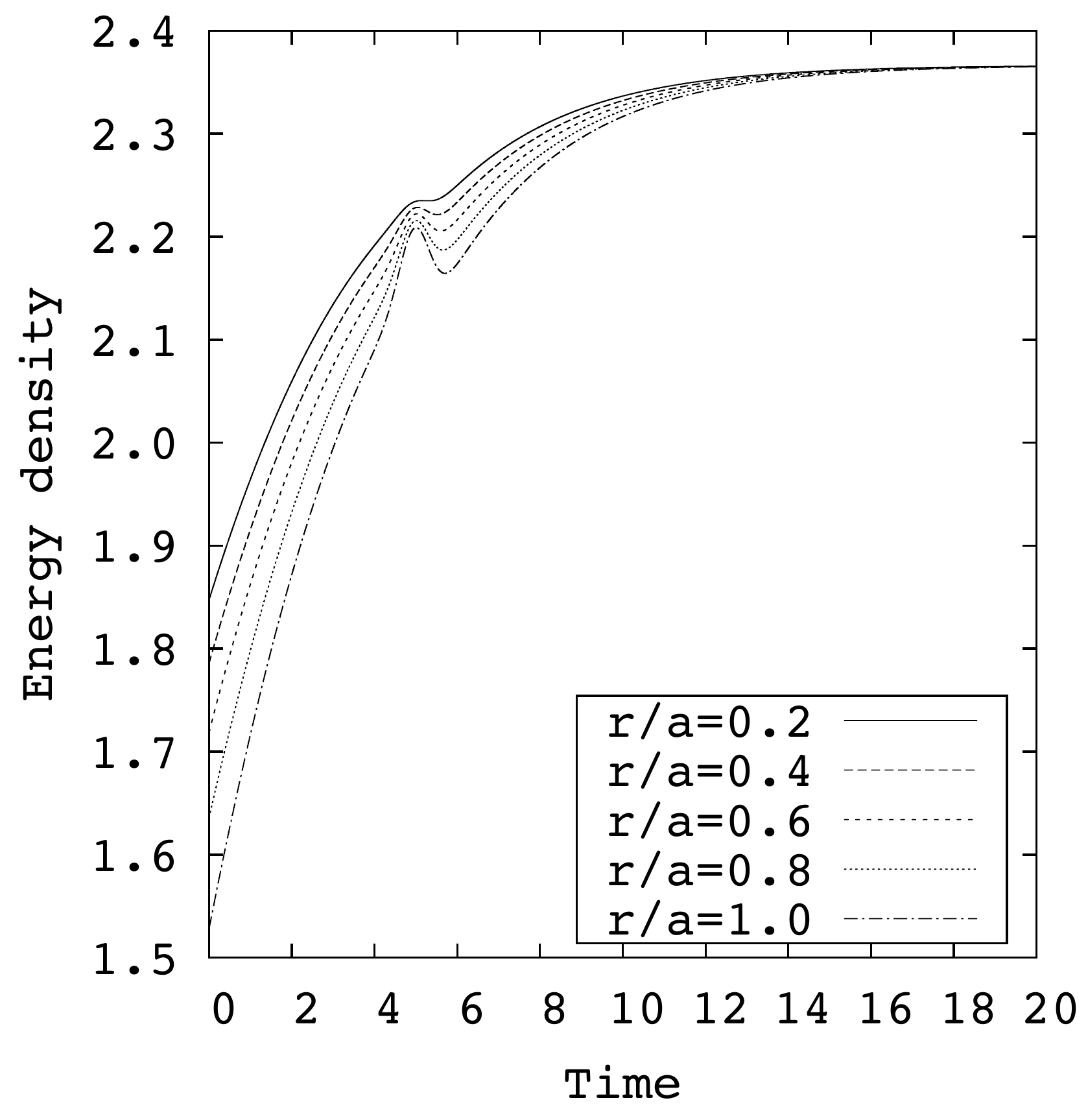}}
\caption{Evolution of the energy density $\rho=\hat\rho+\hat\epsilon$ (multiplied by $10^3$) for the equivalent diffusion limit model (E1) and initial conditions $A(0)=5;\,\,F(0)=0.6;\,\, \Omega(0)=0.8333$.}
\end{center}
\label{fig:dens}
\end{figure}
\begin{figure}[htbp!]
\begin{center}
\scalebox{0.45}{\includegraphics[angle=0]{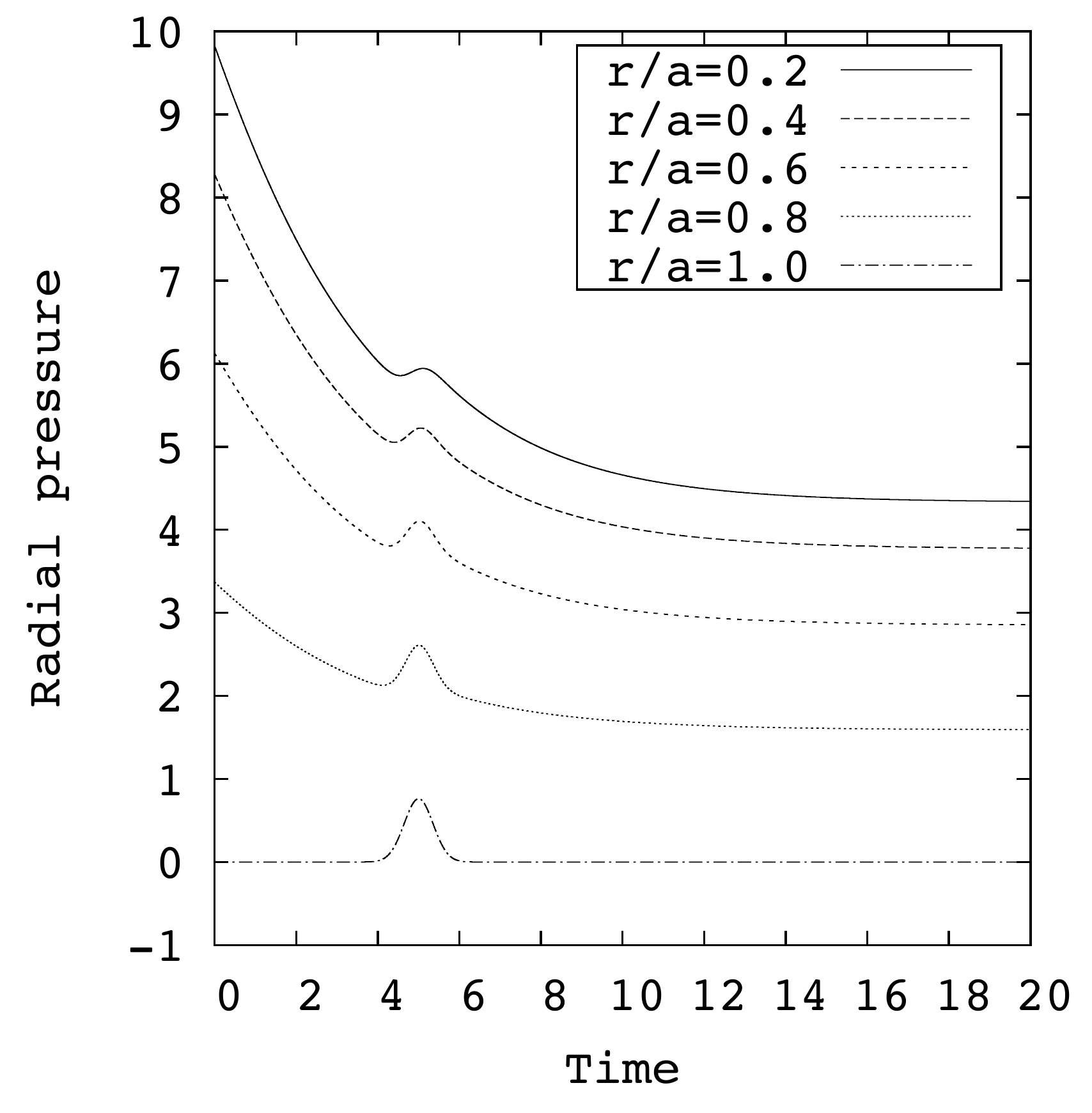}}
\caption{Evolution of the radial pressure $p_r=\hat p + \hat\epsilon$ (multiplied by $10^4$) for the equivalent diffusion limit model (E1) and initial conditions $A(0)=5;\,\,F(0)=0.6;\,\, \Omega(0)=0.8333$.}
\end{center}
\label{fig:pres_dl}
\end{figure}

\begin{figure}[htbp!]
\begin{center}
\scalebox{0.45}{\includegraphics[angle=0]{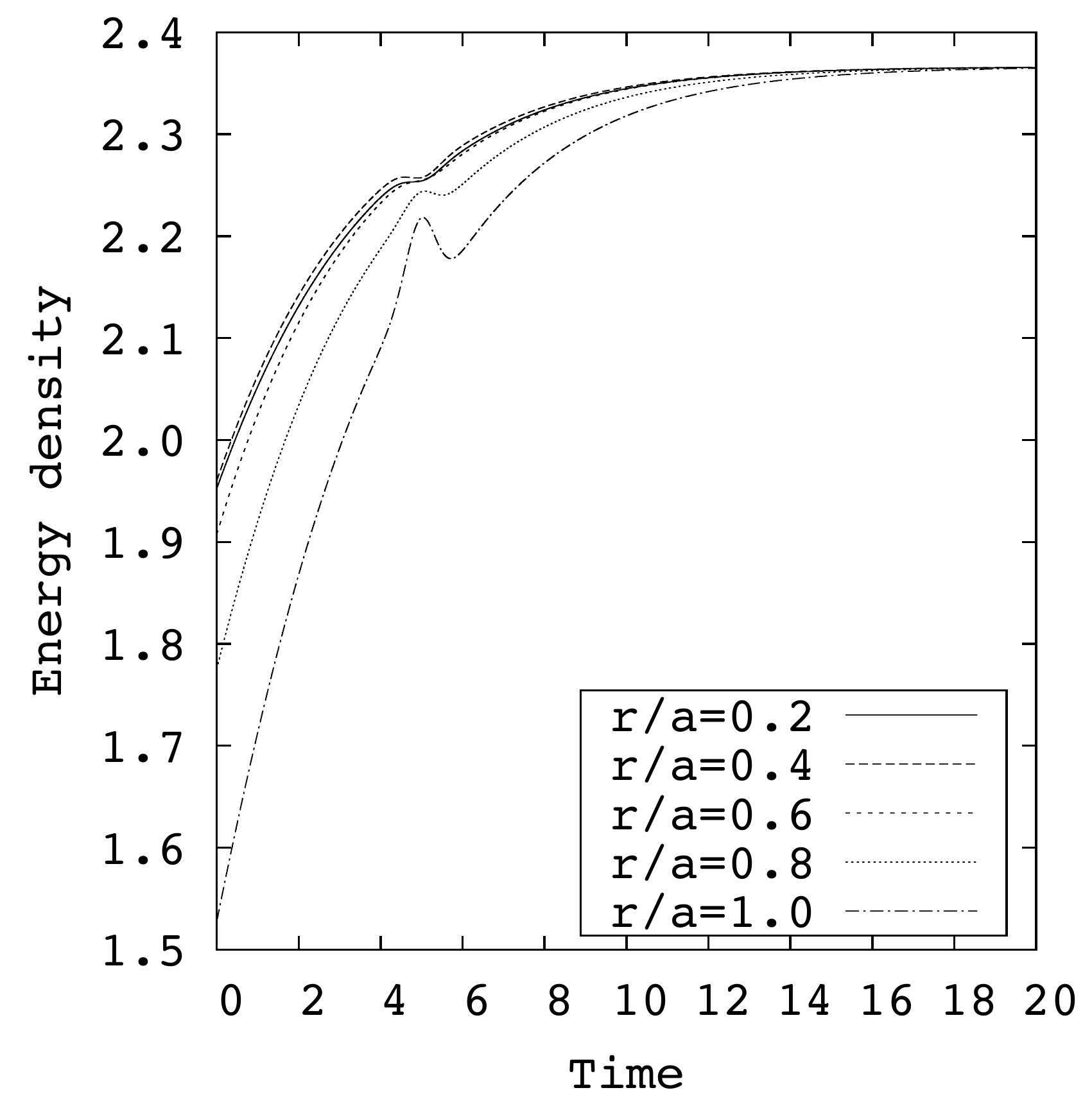}}
\caption{Evolution of the energy density $\rho$ (multiplied by $10^3$) for the diffusion limit model (M4+) with $h=1.2$ and initial conditions $A(0)=5;\,\,F(0)=0.6;\,\, \Omega(0)=0.8333$.}
\end{center}
\label{fig:figure8}
\end{figure}

\begin{figure}[htbp!]
\begin{center}
\scalebox{0.45}{\includegraphics[angle=0]{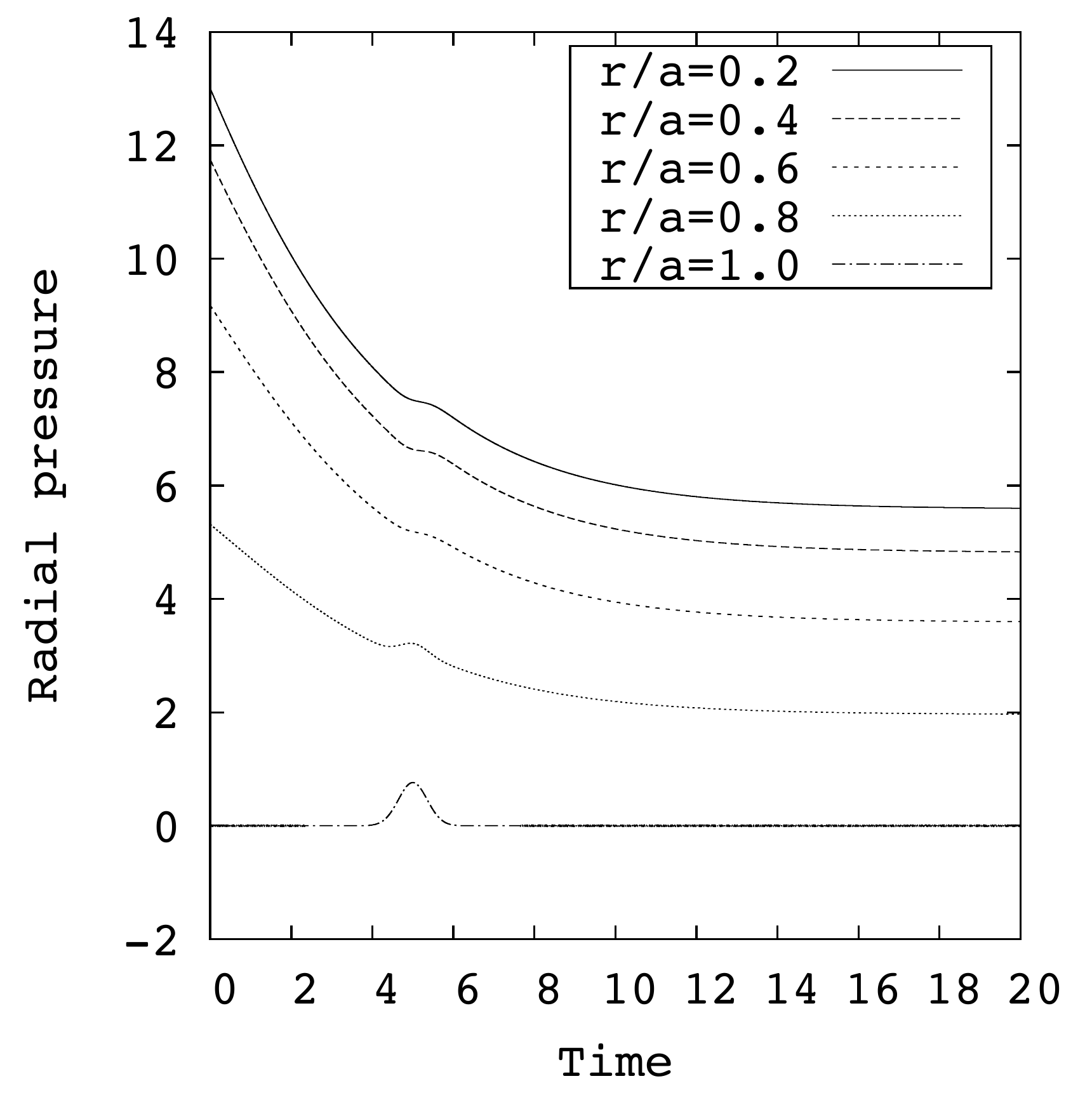}}
\caption{Evolution of the radial pressure $p_r$ (multiplied by $10^4$) for the diffusion limit model (M4+) with $h=1.2$ and initial conditions $A(0)=5;\,\,F(0)=0.6;\,\, \Omega(0)=0.8333$.}
\end{center}
\label{fig:figure9}
\end{figure}

\begin{figure}[htbp!]
\begin{center}
\scalebox{0.45}{\includegraphics[angle=0]{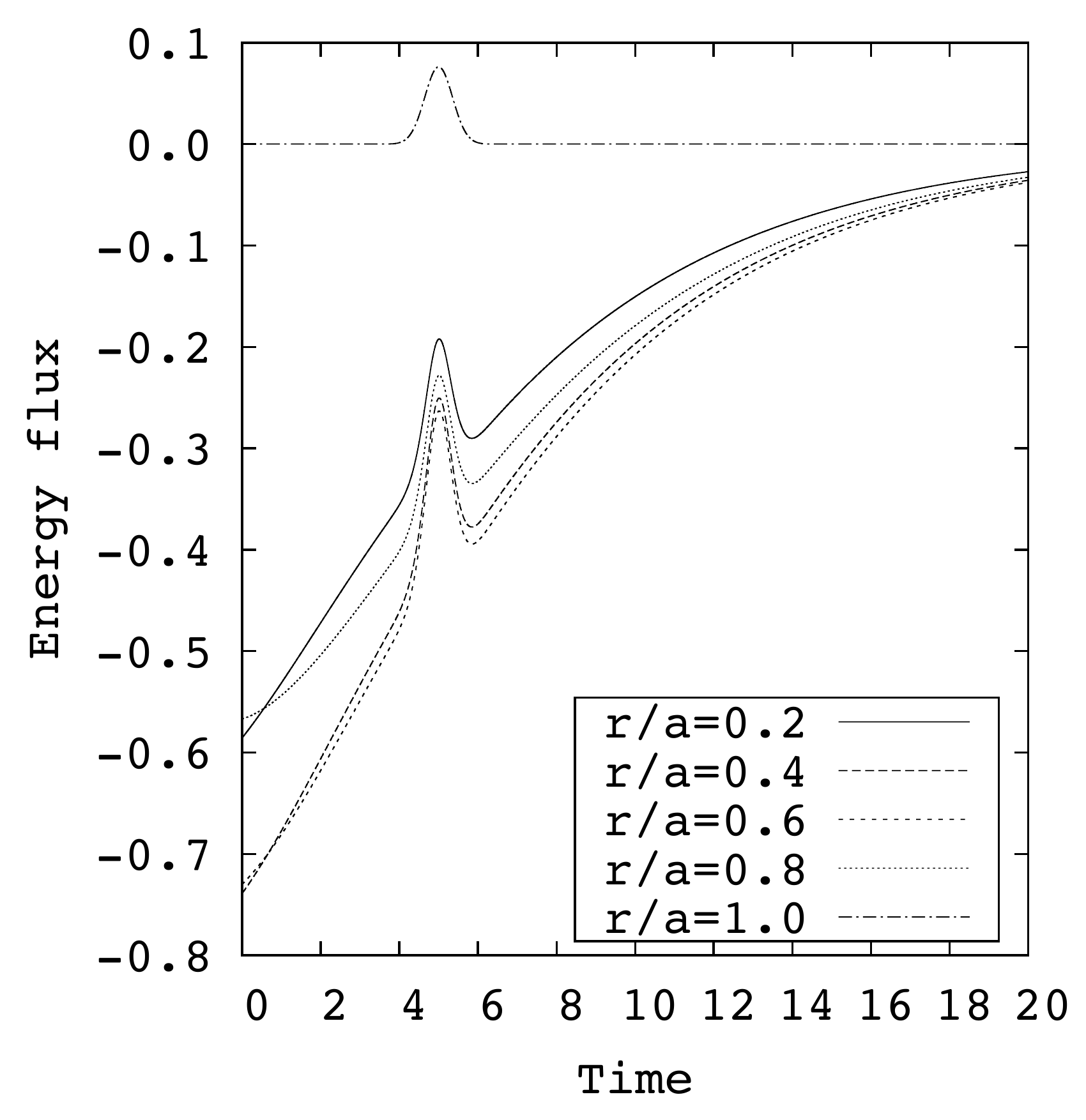}}
\caption{Evolution of the heat flow $\hat q$ (multiplied by $10^3$) for the diffusion limit model (M4+) with $h=1.2$ and initial conditions $A(0)=5;\,\,F(0)=0.6;\,\, \Omega(0)=0.8333$.}
\end{center}
\label{fig:figure10}
\end{figure}

\begin{figure}[htbp!]
\begin{center}
\scalebox{0.45}{\includegraphics[angle=0]{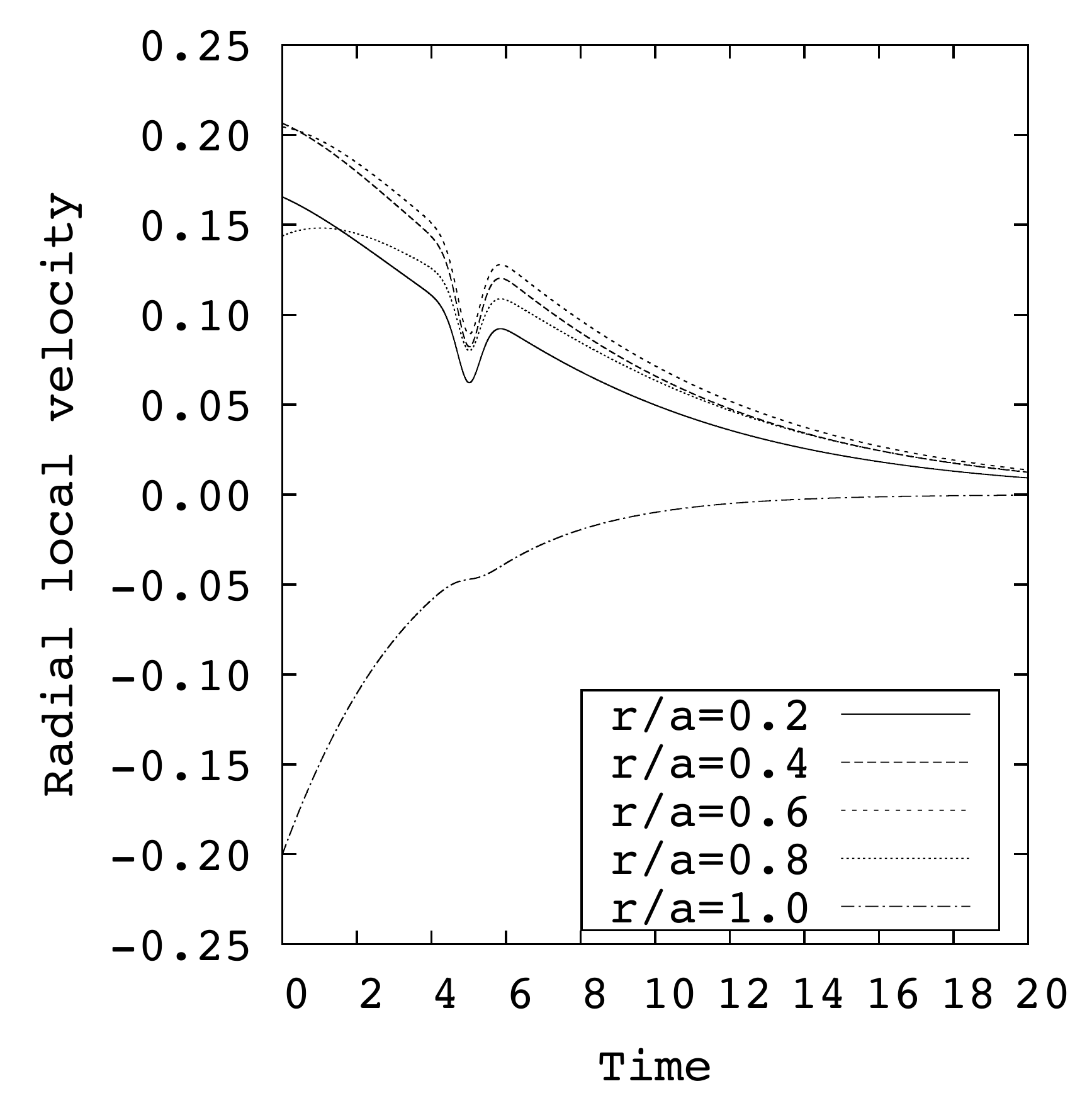}}
\caption{Evolution of the local radial velocity $\omega$ for the diffusion limit model (M4+) with $h=1.2$ and initial conditions $A(0)=5;\,\,F(0)=0.6;\,\, \Omega(0)=0.8333$.}
\end{center}
\label{fig:figure11}
\end{figure}

\begin{figure}[htbp!]
\begin{center}
\scalebox{0.45}{\includegraphics[angle=0]{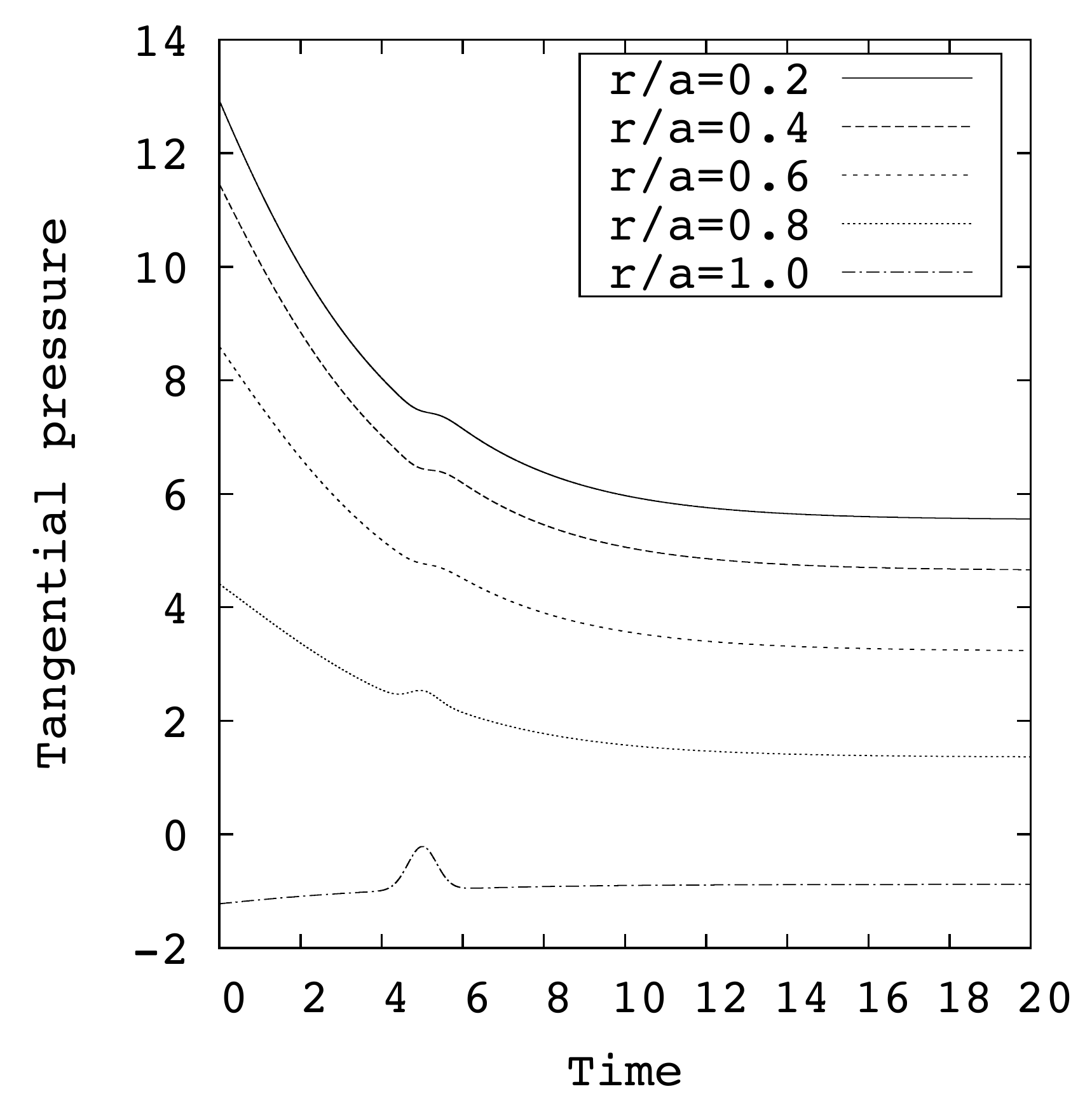}}
\caption{Evolution of the tangential pressure $p_t$ (multiplied by $10^4$) for the diffusion limit model (M4+) with $h=1.2$ and initial conditions $A(0)=5;\,\,F(0)=0.6;\,\, \Omega(0)=0.8333$.}
\end{center}
\label{fig:figure12}
\end{figure}

\begin{figure}[htbp!]
\begin{center}
\scalebox{0.45}{\includegraphics[angle=0]{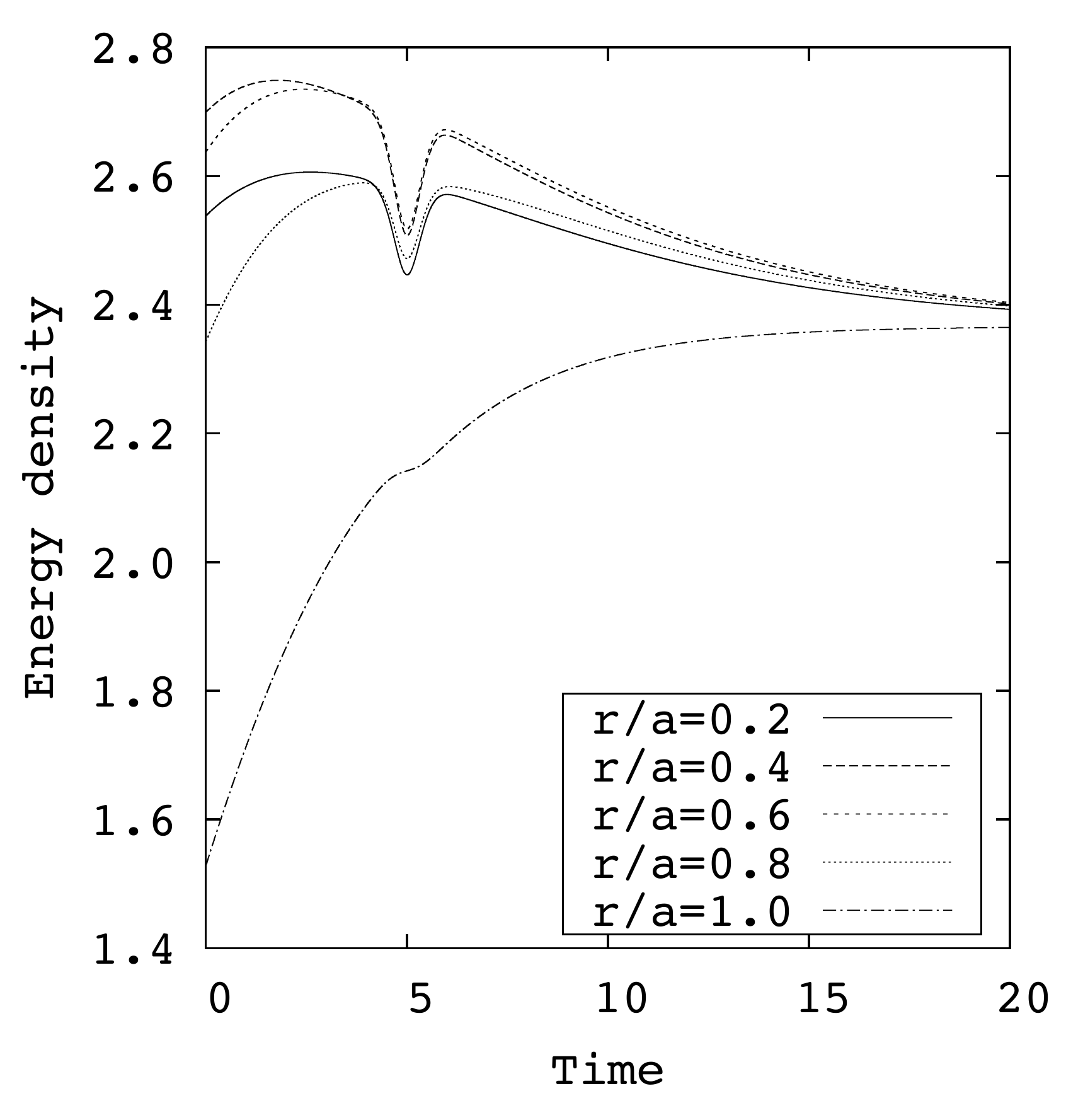}}
\caption{Evolution of the energy density $\hat \rho=\rho-\hat q$ (multiplied by $10^3$) for the diffusion limit model (E2) with $h=1.2$ and initial conditions $A(0)=5;\,\,F(0)=0.6;\,\, \Omega(0)=0.8333$.}
\end{center}
\label{fig:figure13}
\end{figure}

\begin{figure}[htbp!]
\begin{center}
\scalebox{0.45}{\includegraphics[angle=0]{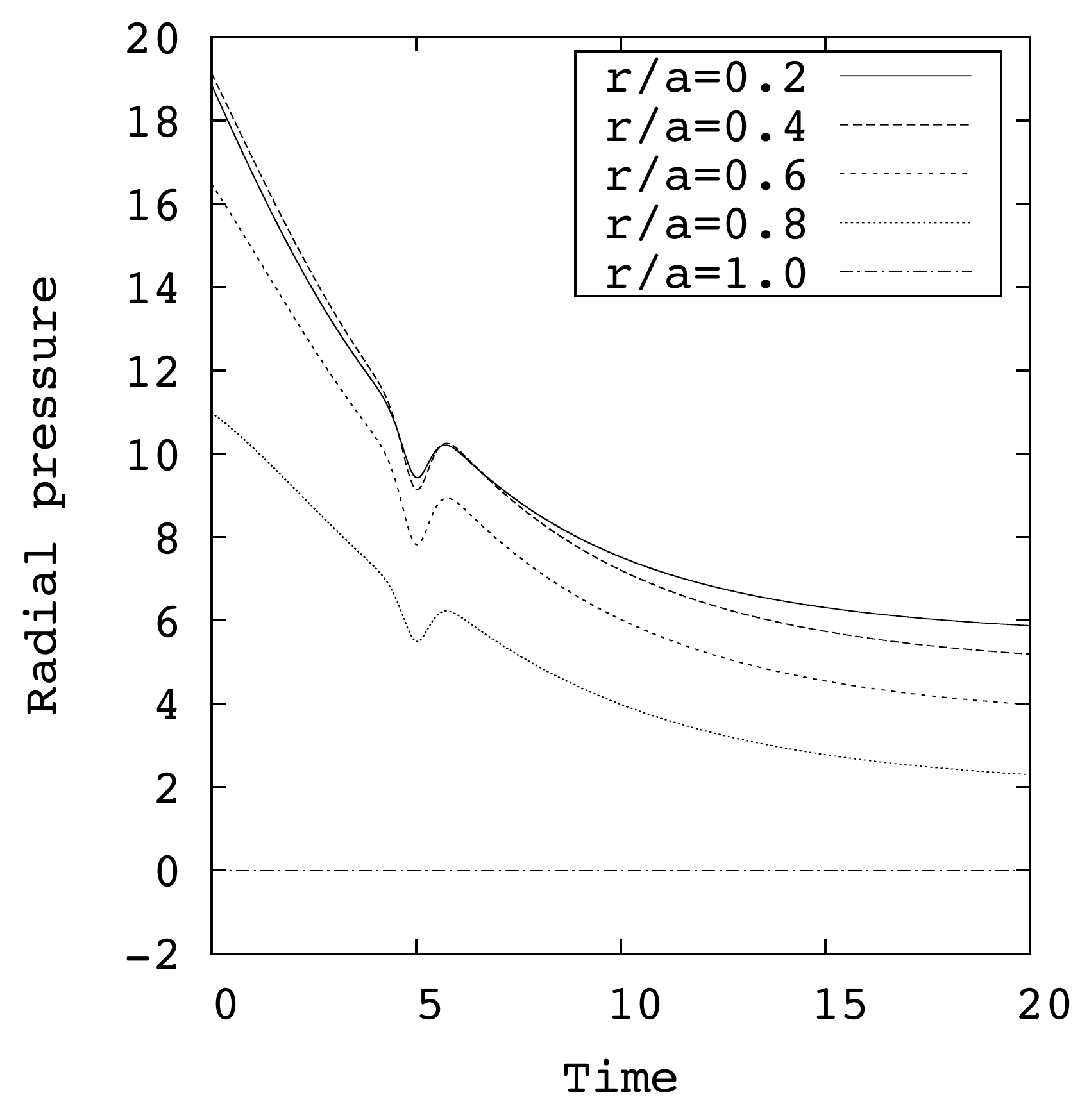}}
\caption{Evolution of the radial pressure $\hat p=p_r - \hat q$ (multiplied by $10^4$) for the diffusion limit model (E2) with $h=1.2$ and initial conditions $A(0)=5;\,\,F(0)=0.6;\,\, \Omega(0)=0.8333$.}
\end{center}
\label{fig:figure14}
\end{figure}
With the following set of initial conditions we can integrate numerically
the system of equations at the surface and go inside the distribution without restrictions:
$$A(0)=5;\,\,\, F(0)=0.6;\,\,\, \Omega(0)=0.8333$$
with $L$ given as a Gaussian pulse carrying away a fraction of total initial mass ($\approx 1\%$): 
$$L=L_0e^{\displaystyle{-(\tau-\tau_0)^2/\sigma^2}},$$
where $\sigma=4$ and $\tau_0=5$.
Figure 1 displays a typical situation in which we can see how the generic fluid behaves
depending on anisotropy and transport mechanism. {Although the behavior shown in figure 1 for each model has been reported in previous investigations \cite{hjr80}, \cite{hje87}, \cite{chew82}, \cite{br92}, for the sake of completeness and to gain physical intuition we summarize and explain these results here. It is clear from Eq. (\ref{TOV}) that anisotropy, second term of the RHS, increases (or decreases) the effective gravitation depending on the tangential pressure and irrespective of the transport mechanism \cite{chew82}. On the other hand, a positive heat flow always diminishes the effective gravitation irrespective of the anisotropy \cite{hje87}. For these reasons the radius for model M4+ ($p_t<p_r$) is smaller than that of model M4$-$ ($p_t>p_r$); model M1 ($p_t=p_r$) is somewhere in between.}
Figures 2--5 display results for model M1 and figures 6--7 for the equivalent model E1. Note that we get $\rho$ and $p_r$ of model E1 from $\hat\rho$, $\hat p$ and $\hat\epsilon$ of model M1; all the other physical variables ($q$, $\omega$ and $p_t=\hat p$) are undistinguishable for these two equivalent models. 
Finally, making numerical experiments we find that model M4+ behaves at the surface like model M1 when $h=1.2$, which is shown in figures 8--12. The equivalent model E2 is displayed in figures 13 and 14.
Observe how physical variables in model E1 (figures 6 and 7) behave like physical variables in model M4+ (figures 8 and 9)
and how physical variables in model E2 (figures 13 and 14) behaves like model M1 (figures 2 and 3). Except for the tangential pressure, the isotropic fluid in the free streaming limit can be equivalent to the anisotropic fluid in the diffusion limit. The GEoS 
works remarkably well to illustrate the equivalence, {displaying only quantitative differences for models E1 and M4+; E2 and M1. In this sense, the two equivalent models E1 and E2 are qualitatively the same as M4+ and M1, respectively}. It suggests that if we impose $[p_t]_a=0$
the performance of the GEoS will get better, but this extra condition is out of our master model.

We want to close this section with the following comment. Setting up the system we realized that for a compact configuration, initially at rest, the fraction of radiated mass
was bounded in the diffusion approximation. Otherwise the local radial velocity was out of the real domain. This situation changes radically with a greater initial velocity of collapse or a lesser fraction of radiated mass (less than $0.01\%$). 
For the free streaming this restriction does not exist, allowing to radiate of more than a $10\%$ of the initial total mass.
This ``asymmetry'' is not in favor of general equivalence but
this was obviously overcome. 

\section{Concluding remarks}
{We have shown how radiating fluids can be equivalent,
when the interior transport mechanisms are opposite and
extreme, namely, for the free streaming and the diffusion
approximations. {From the formal point of view the equivalence between nonadiabatic fluids is indistinguishable when applied to an isotropic fluid with free streaming and to a diffusive anisotropic fluid. One can be derived from the other.} In order to show {numerically} this equivalence, without loss of generality, 
we have used the postquasistatic approach to consider
fluids leaving equilibrium.}
The equivalence can be total
or partial, depending on the initial--boundary conditions and the equation of state.
We illustrate by means of a master model how an isotropic fluid with free streaming can be equivalent to an anisotropic with heat flow dissipation. They are mathematically and physically
equivalent  from each other when using the equation of state (\ref{ieos}). In a generic situation the equivalence depends strongly on the equation of state, initial conditions, junction conditions and exterior spacetime.
We have used in the past the GEoS (\ref{CEO}) to model different physical scenarios.
Viscous \cite{prrb10} or electrically charged \cite{rbrp10} fluids can be seen as anisotropic, but they are not equivalent to isotropic fluids with free streaming or anisotropic fluid with heat flow.

{When coherent radiation is important (scalar, electromagnetic or gravitational radiations) the associated fluids induce
anisotropy and heat flow but the observers are always
resting at infinity.}

Within the family of fluids than can be equivalent it is then plausible that {\it what anisotropy gives (stiffening) heat flow takes (softening)}. In fact, in this case the exterior spacetime is the same from the geometrical point of view. But a judicious observer realizes that,
in general, dissipation in the streaming out limit can never be as that of heat flow.
Mathematically, the reason can be easily understood. In the streaming out limit the algebraic equation to get the radial local velocity from the field equations is a linear expression, while in the diffusion limit the analogous equation is quadratic. In consequence the associated discriminant imposes a severe restriction
on the initial velocity and the fraction of total mass that can be radiated away.
We have shown that the only exception emerges when the two equivalent fluids
are linked by the particular equation of state (\ref{ieos}). 
{Apparently the diffusion mechanism does not accept static
equilibrium. A complex velocity is as unacceptable as a negative energy density or
a superluminous local radial velocity. Therefore a complex velocity is not
physically reliable.
Interestingly, in comoving coordinates that asymmetry is not
present anymore \cite{hb10}.}

{The fundamental character that we can ascribe to two--fluids
is out of proportion, they are simply dissipative fluids.}
Physically: i) The mean free path for both limits are very different; ii) Temperature profiles from a transport equation, as in the M\"uller--Israel--Stewart theory \cite{Muller67}--\cite{131}, \cite{hjr9}, \cite{HJ}, makes the complete equivalence not viable at all, enhancing asymmetry. Even so, the master fluid idea is 
attractive, isotropic with free streaming or anisotropic with heat flow, to model a number of situations; in between we can use the Eddington factor \cite{ans05}. Thus, equivalence in this context can be used as a first approximation to study dissipative fluids.

Our argument in this paper was developed clearly thanks to a known approach \cite{b64}, to treat matter from a comoving frame with the fluid in the local Minkowskian frame, called Bondian \cite{b09} (usually referred as global or Eulerian). We know that the same conclusion can arise from Misner--Sharp's comoving coordinates \cite{ms64} (local or Lagrangian frame). In this sense, 
we get an unified treatment of nonadiabatic fluids.
\acknowledgments
Thanks to Luis Herrera, Carlos Peralta, Luis Rosales and Beltr\'an Rodr\'\i guez--Mueller for reading, criticism and interesting comments.

\end{document}